\documentstyle[preprint,prd,aps,floats,epsf,tighten,amsmath]{revtex}

\begin{document}

\preprint{\vbox{
\hbox{JLAB-THY-01-13}
\hbox{FSU-CSIT-01-24}
\hbox{hep-lat/0105004}
}}

\title{Are Topological Charge Fluctuations in QCD Instanton Dominated?}
\author{Robert G. Edwards}
\address{
Jefferson Lab,
12000 Jefferson Avenue,
MS 12H2,
Newport News, VA 23606, USA}
\author{Urs M. Heller}
\address{
CSIT, Florida State University, 
Tallahassee, FL 32306-4120, USA}

\date{\today}

\maketitle

\begin{abstract} 
We consider a recent proposal by Horv\'ath {\em et al.} to address the
question whether topological charge fluctuations in QCD are instanton
dominated via the response of fermions using lattice fermions with exact
chiral symmetry, the overlap fermions. Considering several volumes and
lattice spacings we find strong evidence for chirality of a finite
density of low-lying eigenvectors of the overlap-Dirac operator in the
regions where these modes are peaked. This result suggests instanton
dominance of topological charge fluctuations in quenched QCD.
\end{abstract}

\section{Introduction}

Understanding the mechanisms of confinement and chiral symmetry breaking
in QCD has been a major goal of non-perturbative field theory for many
years. A good candidate for the source of chiral symmetry breaking are
topological fluctuations of the gauge fields associated with instantons
and anti-instantons. Through mixing, the ``would be zero modes'' from
the instantons and anti-instantons become the small modes that lead,
when present with a finite density, to chiral symmetry breaking. Many
of the low energy properties of QCD can be explained phenomenologically
by the interactions of the fermions with instantons and anti-instantons
\cite{inst_liquid,diak}. To our knowledge, confinement, though, is not one
of those properties.

However, as pointed out by Witten~\cite{Witten}, there is an inconsistency
between instanton based phenomenology and large-$N_c$ QCD. Instantons
would produce an $\eta^\prime$ mass that vanishes exponentially for
large $N_c$, while considerations based on large-$N_c$ chiral dynamics
suggest that the $\eta^\prime$ mass should be of order $1/N_c$. The
topological charge fluctuations then should be associated not to instantons
but rather some other, confinement-related vacuum fluctuations. And the
strong attraction produced by these confinement-related fluctuations
would also induce the breaking of chiral symmetry.

In an attempt to settle the issue, whether topological charge fluctuations
are instanton dominated or not, Horv\'ath, Isgur, McCune and
Thacker~\cite{HIMT} investigated the response of fermions by considering
low lying eigenmodes of a lattice discretized Dirac operator and
quantifying the extend to which, in the peak regions of these eigenmodes
they are chiral, as would be expected if the almost zero modes around
instantons dominate these low lying eigenmodes. They introduced a local
``chiral orientation'' parameter, or ``chirality'' parameter, $X(x)$,
via the definition
\begin{equation}
\tan \Bigl( \frac{\pi}{4} (1+X(x)) \Bigr) = \biggl(
 \frac{\psi^\dagger_L(x) \psi_L(x)} {\psi^\dagger_R(x) \psi_R(x)}
 \biggr)^{1/2} ~,
\label{eq:chir_1}
\end{equation}
where $\psi_L(x)$ and $\psi_R(x)$ are the left- and right-handed components
of the eigenmode. Chirality near $\pm 1$ would be obtained from almost
zero modes due to instantons and anti-instantons, lifted from being exact
zero modes by their mixing.

Horv\'ath {\em et al.}~\cite{HIMT} chose the Wilson discretization for their
lattice Dirac operator and claim to have found evidence against instanton
dominance of topological charge fluctuations from the behavior of the
chirality parameter. However, the Wilson-Dirac operator explicitly
breaks chiral symmetry at any finite lattice spacing $a$. As a consequence,
the Wilson-Dirac operator does not have exact (chiral) zero modes in
gauge fields with a non-trivial topological charge. And the discretization
effects responsible might well also strongly affect the chirality
parameter. To our knowledge, the magnitude of such distortion effects
at the presently accessible lattice spacings are not known. However,
the scaling violations for hadron masses due to the ${\cal O}(a)$ lattice
artifacts at the lattice spacing used by Horv\'ath {\em et al.~}($a \simeq 0.17$
fm at $\beta=5.7$) are known to be as large as 30--40\%~\cite{EHK_clov}.

In a response to \cite{HIMT} DeGrand and Hasenfratz presented evidence that
is suggestive of instanton dominance of topological charge
fluctuations~\cite{DH}. They used a variant of the overlap-Dirac operator,
proposed and investigated in~\cite{TD_ov}, for the lattice Dirac operator.
The overlap-Dirac operator~\cite{Herbert}, and any variant that substitutes
the Wilson-Dirac operator used in its construction by another suitable
Wilson-like discretization of the Dirac operator, has the chiral
and topological properties of the continuum Dirac operator even at finite
lattice spacing, {\it e.g.} exact chiral zero modes in gauge fields with
a non-trivial topological charge. However, DeGrand and Hasenfratz coupled
their overlap fermions not directly to the ``rough'' lattice gauge fields,
but rather to ``APE smeared'' gauge fields, essentially Gaussian smeared
fields with fixed width in lattice units. While such a fermion action is
local and therefore does not change universality class, the distortions
the smearing procedure might induce at finite lattice spacings are not
known. Indeed, the level of smearing used by DeGrand and Hasenfratz removes
ultraviolet fluctuations in the gauge field to an extent where pure gauge
observables start to ``feel'' and be able to identify instanton-like
topological excitations that on the ``rough'' lattice gauge fields are
completely obscured by the dominant ultraviolet fluctuations. For example,
the naive topological charge measurements gives results that are clearly
peaked around integer values. The smearing used distorts the short
distance part of the heavy quark potential, to a distance of about 3
lattice spacings, and lowers the value obtained for $f_B$ when the fat
links are used in the clover Dirac operator~\cite{MILC_Pisa}. Whether
the smearing affects properties of light fermions has, to our knowledge,
not been investigated thoroughly.

In an earlier paper, using the same variant of the overlap-Dirac operator,
DeGrand and Hasenfratz produced evidence that the chiral density
$\omega(x) = \psi^\dagger \gamma_5 \psi(x)$ comes in lumps on the lattice
\cite{DH_long}, and furthermore that the correlation function of the
chiral density with itself is similar in shape to the correlation
function of the chiral density $\omega(x)$ with the topological charge
density $Q(x)$ obtained from an APE-smeared operator. But, since APE-smearing
is involved, these results are not beyond doubt. Indeed, the authors
of Ref.~\cite{HIMT} argue that the ``lumpiness'' observed in the correlation
function of the chiral density $\omega(x)$ could occur without these lumps
being {\it purely} right-handed or left-handed.

Here we address these questions using the standard overlap-Dirac
operator~\cite{Herbert} with the fermions coupled to the original lattice
gauge fields. We thereby avoid potential distortions due to smearing while
we retain all chiral properties including existence of exact zero modes and
thus avoid the lattice artifacts, unknown in magnitude and significance, of
Wilson fermions.

\section{Setup}

We consider the standard massless overlap-Dirac operator~\cite{Herbert}
\begin{equation}
D(0) = \frac{1}{2} \bigl[ 1 + \gamma_5 \epsilon(H_w(M)) \bigr]
\label{eq:Dov}
\end{equation}
with $H_w(M) = \gamma D_w(-M)$ and $D_w(M)$ the usual Wilson-Dirac operator.
Then, $H^2(0) = D^\dagger(0) D(0)$ commutes with $\gamma_5$ and can
be simultaneously diagonalized~\cite{EHN_practical,ParalComp}. $H^2(0)$
can have zero eigenvalues with chiral eigenmodes, which are also
eigenmodes of $D(0)$, due to global topology. The non-zero eigenmodes of
$H^2(0)$ are doubly degenerate and have opposite chirality,
\begin{equation}
H^2(0) \psi_{\uparrow,\downarrow} = \lambda^2 \psi_{\uparrow,\downarrow}
 ~,\qquad \gamma_5 \psi_\uparrow = \psi_\uparrow ~,
 \gamma_5 \psi_\downarrow = - \psi_\downarrow ~,
\end{equation}
with $0 < \lambda \le 1$.
The non-zero (right) eigenmodes of $D(0)$ are then easily obtained as
\begin{equation}
\psi_\pm = \frac{1}{\sqrt{2}} \Bigl( \psi_\uparrow
 \pm i \psi_\downarrow \Bigr)
\end{equation}
with eigenvalues
\begin{equation}
\lambda_\pm = \lambda^2 \pm i \lambda \sqrt{1 - \lambda^2} ~.
\end{equation}
The ``chirality'' parameter of eq.~(\ref{eq:chir_1}) is then given by
\begin{equation}
\tan \Bigl( \frac{\pi}{4} (1+X(x)) \Bigr) = \biggl(
 \frac{\psi^\dagger_\downarrow(x) \psi_\downarrow(x)}
 {\psi^\dagger_\uparrow(x) \psi_\uparrow(x)} \biggr)^{1/2} ~,
\label{eq:chir_2}
\end{equation}
and is equal for the two related modes $\psi_\pm$. In the following they
will be therefore counted as one mode.
The exact zero modes, of course, have $X(x) \equiv \pm 1$.

We computed low-lying eigenmodes $\psi_{\uparrow,\downarrow}$ of $H^2(0)$
with the Ritz functional algorithm of Ref.~\cite{Ritz}. For the sign
function $\epsilon(H_W)$ in eq.~(\ref{eq:Dov}) we used the optimal
rational approximation of \cite{EHN_practical,ParalComp} with ``projection
of low-lying eigenvectors of $H_w$'' to ensure sufficient accuracy.

\section{Results}

We computed the 20 eigenmodes $\psi_{\uparrow,\downarrow}$ of $H^2(0)$
with smallest $\lambda$ on ensembles of $8^3 \times 16$ pure gauge Wilson
action configurations with $\beta=5.7$ and 5.85, and Wilson-Dirac mass
of $M=1.65$. The gauge configurations have lattice spacing
of about 0.17 and 0.125 fm, respectively, and hence volumes of about 7
and 2 fm$^4$. We also computed the 20 lowest eigenmodes on an ensemble
of $6^3 \times 12$ lattices with $\beta=5.7$, with a volume of about
2.2 fm$^4$, fairly close to the volume of the $\beta=5.85$ ensemble.

\begin{figure}
\centerline{{\setlength{\epsfxsize}{6in}\epsfbox[50 50 626 626]
 {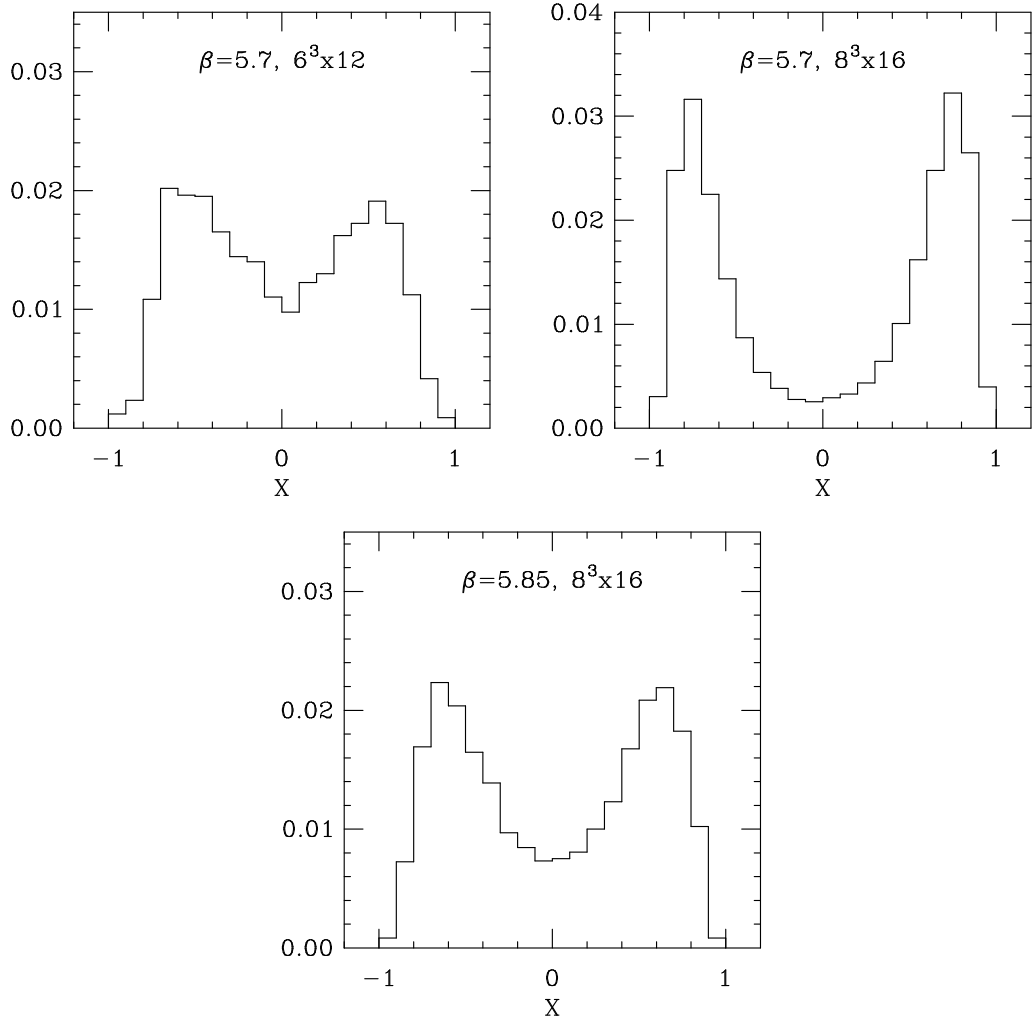}}}
\caption{Chirality histograms for the lowest two non-zero modes of the
overlap-Dirac operator at the 2.5\% sites with the largest
$\psi^\dagger \psi(x)$ on the three ensembles with Wilson gauge action.}
\label{fig:hist_W_2l_2p5pc}
\end{figure}

In Fig.~\ref{fig:hist_W_2l_2p5pc} we show the chirality histograms for the
lowest two non-zero modes at the 2.5\% of the lattice sites with the largest
$\psi^\dagger \psi(x)$. In this, and in all similar figures, the histograms
are normalized such that the area under the histogram is equal to the fraction
of sites considered. In their peaks, the lowest two eigenmodes are fairly
chiral. Comparing the two ensembles with the same gauge coupling (and hence
lattice spacing) we see a dramatic dependence on the physical volume.
In an instanton liquid picture of the vacuum, the number of instantons and
anti-instantons, and hence the number of almost zero modes grows linearly
with the volume. Then it is not surprising that the lowest modes become
increasingly chiral in their peaks with increasing volume. Comparing the
two ensembles with almost equal volume in physical units, we notice a
tendency of the peaks to become more chiral as the lattice spacing is
decreased.

Comparing with Fig.~1 of Ref.~\cite{HIMT} we see that in their peaks the
lowest two non-zero eigenmodes of the overlap-Dirac operator appear to be
more chiral than the real modes of the Wilson-Dirac operator. Considering
that Ref.~\cite{HIMT} used a lattice with larger volume of about 35 fm$^4$
and the dramatic volume dependence of the chirality histogram observed
here leads one to suspect that the chiral symmetry breaking lattice
artifacts of the Wilson-Dirac operator are truly significant at the
lattice spacing used.

Comparing with Fig.~1 of Ref.~\cite{DH}, on the other hand, where a physical
volume of about 3.5 fm$^4$ was used, indicates that the APE smeared fields
used by DeGrand and Hasenfratz enhance the chirality in the peaks of the
lowest-lying non-zero modes somewhat. DeGrand and Hasenfratz already noted
a slight dependence on the smearing of gauge fields (see Fig.~4 of~\cite{DH}),
though the effect appears much smaller than the effects of the lattice
artifacts from using Wilson fermions.

\begin{figure}
\centerline{{\setlength{\epsfysize}{3in}\epsfbox[250 200 376 526]
 {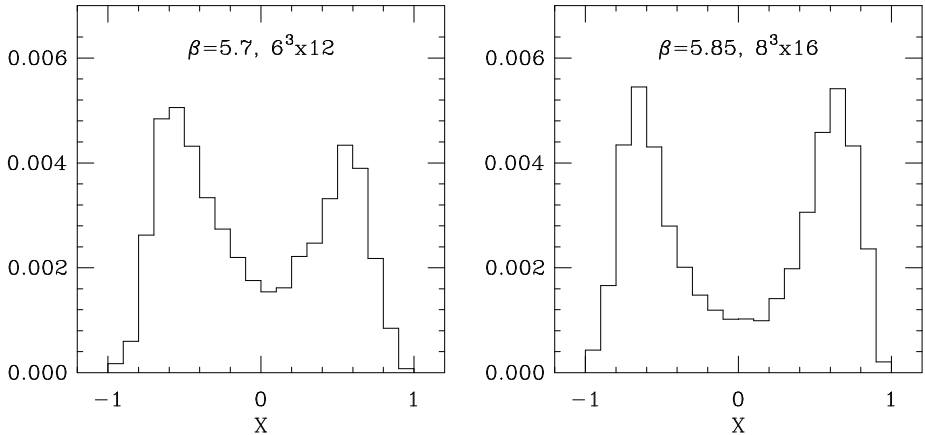}}}
\caption{Chirality histogram similar to Fig.~\protect\ref{fig:hist_W_2l_2p5pc},
but for the lowest four non-zero modes at the 0.5\% sites with the largest
$\psi^\dagger \psi(x)$.}
\label{fig:hist_W_4l_0p5pc}
\end{figure}

\begin{figure}
\centerline{{\setlength{\epsfxsize}{4in}\epsfbox[0 0 576 576]
 {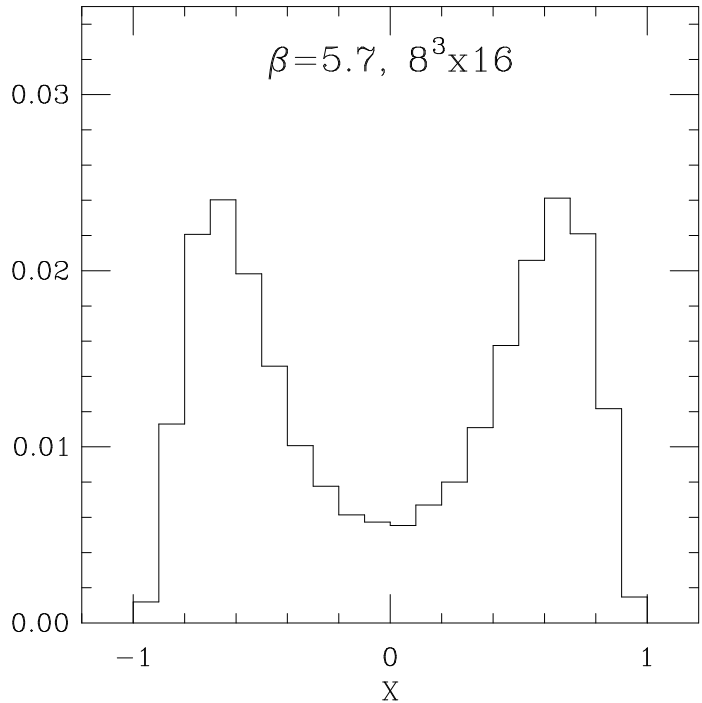}}}
\caption{Chirality histogram for the lowest six non-zero modes at the
2.5\% sites with the largest $\psi^\dagger \psi(x)$ on the lattice at
$\beta=5.7$ with the larger volume.}
\label{fig:hist_570L_6l_2p5pc}
\end{figure}

Considering also the next two higher modes while further restricting the peak
region gives the chirality histograms for the two smaller lattices shown
in Fig.~\ref{fig:hist_W_4l_0p5pc}. These modes appear still rather chiral
in their peaks. On the larger volume, as expected, a larger number of modes
are chiral in their peak, as seen in Fig.~\ref{fig:hist_570L_6l_2p5pc}.

To confirm the trends with volume and lattice spacing, we would like
to consider a still larger volume, as well as a lattice at smaller
lattice spacing. Unfortunately, the numerical implementation of
overlap fermions becomes increasingly difficult and costly, due to the
finite density of modes with almost zero eigenvalues of the underlying
Wilson-Dirac operator~\cite{EHN_rho0}. It is believed that these modes
are caused by lattice artifact dislocations in the gauge fields
generated with Wilson like gauge actions. 
It is known~\cite{DWF_CPPACS,DWF_eigs,DWF_BNL} that these same small
modes are responsible for the residual chiral symmetry breaking
observed for domain wall fermions with finite 5-th dimension $L_s$.
It was found that improving the gauge action via the Symanzik improved 
action suppresses the dislocations and decreases the
density of small modes~\cite{EHN_rho0}.
It has also been observed that the residual
chiral symmetry breaking for domain wall fermions on gauge fields
generated with an improved gauge action is much
reduced~\cite{DWF_eigs,DWF_IW,PV}. And in Ref.~\cite{Liu} it was noted
that the cost of implementing the overlap-Dirac operator on gauge fields
with a tree-level Symanzik improved gauge action is lowered.

We made a study of various gauge actions to find which one
lowered the density of small modes and would consequently lower the
cost of implementing the overlap-Dirac operator. The density of zero
eigenvalues of the hermitian Wilson-Dirac operator $\rho(0)$ at the
mass $M=1.65$ was determined for quenched backgrounds from the
Iwasaki~\cite{IW}, tree-level tadpole-improved and 1-loop tadpole-improved
L\"uscher-Weisz gauge action~\cite{LW}. 
This was done by computing
the low lying eigenvalues of $H_w(M)$ using the Ritz functional~\cite{Ritz}.
The low lying eigenvalues over the whole ensemble are then used to
obtain the integral of the spectral density function,
$\int_0^\lambda \rho(\lambda^\prime)d\lambda^\prime$.
A linear fit in $\lambda$ is made, and $\rho(0)$ is obtained as the 
coefficient of the linear term. 
Comparisons were made with the
Wilson gauge action at two scales determined by the string tension.
The results shown in Table~\ref{tab:rho0} show the quenched Iwasaki
gauge action to have the lowest $\rho(0)$. For example, at the lattice
spacing equivalent to Wilson gauge action $\beta=5.7$ the $\rho(0)$ is
reduced by a factor of about 5. 

We therefore generated gauge field ensembles with Iwasaki's action with
$\beta = 2.2872$ and 2.45, chosen to give the same lattice spacing,
in units of the string tension, as the ensembles already considered
with Wilson action at $\beta=5.7$ and 5.85. As expected, since the density of
small modes of the Wilson-Dirac operator used in the overlap fermion
construction were considerably reduced, the overlap fermion computation
became much faster. For example, at the lattice spacing equivalent to
Wilson $\beta=5.7$ and lattice size $8^3\times 16$ the cost of the inner
conjugate-gradient step in applying the overlap-Dirac operator was reduced
from between 200 to 400 iterations to between 120 to 170 iterations.
The average time per configuration to compute the 20 lowest eigenvectors
was reduced by roughly a factor of 3.
This made it feasible to consider lattices of size
$12^3 \times 16$, giving a volume of about 23.6 fm$^4$ at $\beta=2.2872$
and 6.7 fm$^4$ at $\beta=2.45$, {\it i.e.} about the same volume as the
$8^3 \times 16$ lattice at $\beta=2.2872$. We also generated a
$12^3 \times 16$ at $\beta=2.65$, chosen such that the volume is about
2.1 fm$^4$, the same volume as the smallest lattice at the other two
gauge couplings. The lattice size and volume, both in lattice and physical
units, as well as the average number of zero modes and the fluctuation
of the topological charge, obtained from the number of zero modes, of
all ensembles considered are listed in Table~\ref{tab:Q_Qsq}.

\begin{table}
\caption{Comparison of the density of zero eigenvalues $\rho(0)$ of the 
hermitian Wilson-Dirac operator, in lattice units,
for various gauge actions. The comparisons
were made at the same lattice spacings as the Wilson gauge action $\beta=5.85$
and $5.7$. The scale was set using the string tension. Shown are the results
for the Wilson, Iwasaki, tree-level and 1-loop tadpole-improved 
L\"uscher-Weisz gauge actions. The string tensions for the Wilson gauge
action were taken from Ref.~\protect\cite{EHK_string}. The string tension for
the 1-loop tadpole-improved L\"uscher-Weisz gauge action comes from an
interpolation of previous results, while the others were computed in this work.
All tests used a Wilson-Dirac mass of $M=1.65$.
The Iwasaki gauge action has the lowest $\rho(0)$ at fixed $a\sqrt\sigma$.
}
\label{tab:rho0}
\begin{tabular}{|c||l|r|r|l|}
 action & $\beta$ & $V$ $a^4$ & $a\sqrt\sigma$ & $\rho(0)\times 10^{5}$ \\
\hline
Wilson & 5.70   &  $8^3 \times 16$ &  0.3917 & 162(9) \\
       & 5.85   &  $8^3 \times 16$ &  0.2864 & 21(3)  \\
       & 6.00   & $24^3 \times 32$ &  0.2197 & 2.3(2) \\
\hline
Iwasaki & 2.2872 &  $8^3 \times 16$ &  0.3885(93) & 32(4) \\
        & 2.45   & $12^3 \times 16$ &  0.2840(53) & 2.4(9) \\
\hline
0-loop TI LW & 7.26   &  $8^3 \times 16$ &  0.3900(76) & 58(6) \\
        & 7.60   & $12^3 \times 16$ &  0.2872(39) & 4.3(8) \\
\hline
1-loop TI LW & 7.79   &  $8^3 \times 16$ &  0.3917 & 44(5) \\
        & 8.13   & $12^3 \times 16$ &  0.2864 & 3.9(7) \\
\hline
\end{tabular}
\end{table}

\begin{table}
\caption{The gauge ensembles with Wilson glue in the first three lines and
with Iwasaki glue in the rest. Listed are the volume in lattice units and in
units of fm$^4$, the number of configurations analyzed, the average number
of zero modes and the average of the square of the global topological
charge as obtained from the number of zero modes.}
\label{tab:Q_Qsq}
\begin{tabular}{|l|r|r|l|l|l|}
 $\beta$ & $V$ [$a^4$] & $V$ [fm$^4$] & $N_{cfg}$ & $\langle |Q| \rangle$ &
 $\langle Q^2 \rangle$  \\
\hline
 5.70   &  $6^3 \times 12$ &  2.2 &  50 & 1.2(2) & 2.8(5)  \\
 5.70   &  $8^3 \times 16$ &  7.0 &  48 & 2.0(2) & 7(1)    \\
 5.85   &  $8^3 \times 16$ &  2.0 &  93 & 1.3(1) & 2.6(4)  \\
\hline
 2.2872 &  $6^3 \times 12$ &  2.2 & 100 & 1.0(1) & 1.7(2)  \\
 2.2872 &  $8^3 \times 16$ &  7.0 &  69 & 2.0(2) & 5(1)    \\
 2.2872 & $12^3 \times 16$ & 23.6 &  10 & 4.2(9) & 25(8)   \\
 2.45   &  $8^3 \times 16$ &  2.0 & 197 & 1.0(1) & 1.8(2)  \\
 2.45   & $12^3 \times 16$ &  6.7 &  47 & 2.3(4) & 8(1)    \\
 2.65   & $12^3 \times 16$ &  2.1 &  50 & 1.3(2) & 2.4(6)  \\
\hline
\end{tabular}
\end{table}

\begin{table}
\caption{The average of the sum of $\psi^\dagger \psi(x)$ over the 2.5\%
of sites kept for the chirality histograms as  a percentage of the sum over
all sites for some low-lying eigenvectors for the various gauge field 
ensembles.}
\label{tab:psi_sq}
\begin{tabular}{|l|r|l|l|l|l|l|}
 $\beta$ & $V$ [$a^4$] & ev 1 & ev 2 & ev 4 & ev 6 & ev 20 \\
\hline
 5.70   &  $6^3 \times 12$ &  8.7(3) &  7.9(3) &  7.0(2) &  6.4(2) &  5.1(1) \\
 5.70   &  $8^3 \times 16$ & 11.4(3) & 11.0(3) &  9.7(2) &  9.3(2) &  6.9(5) \\
 5.85   &  $8^3 \times 16$ & 10.6(3) &  9.5(3) &  7.9(2) &  7.5(2) &  6.1(1) \\
\hline
 2.2872 &  $6^3 \times 12$ &  8.9(2) &  7.8(2) &  6.9(2) &  6.3(1) &  5.1(1) \\
 2.2872 &  $8^3 \times 16$ & 11.5(3) & 10.6(2) &  9.3(2) &  8.5(2) &  6.2(2) \\
 2.2872 & $12^3 \times 16$ & 12.9(5) & 13.4(6) & 12.3(4) & 12.0(4) & 11.3(5) \\
 2.45   &  $8^3 \times 16$ &  9.9(2) &  8.7(1) &  7.7(1) &  7.2(1) &  6.0(1) \\
 2.45   & $12^3 \times 16$ & 12.4(3) & 11.5(3) & 10.3(2) &  9.9(3) &  7.8(3) \\
 2.65   & $12^3 \times 16$ & 10.4(3) &  9.2(2) &  8.0(1) &  7.7(1) &  6.9(3) \\
\hline
\end{tabular}
\end{table}

\begin{figure}
\vspace{2cm}
\centerline{{\setlength{\epsfxsize}{7in}\epsfbox[0 0 676 676]
 {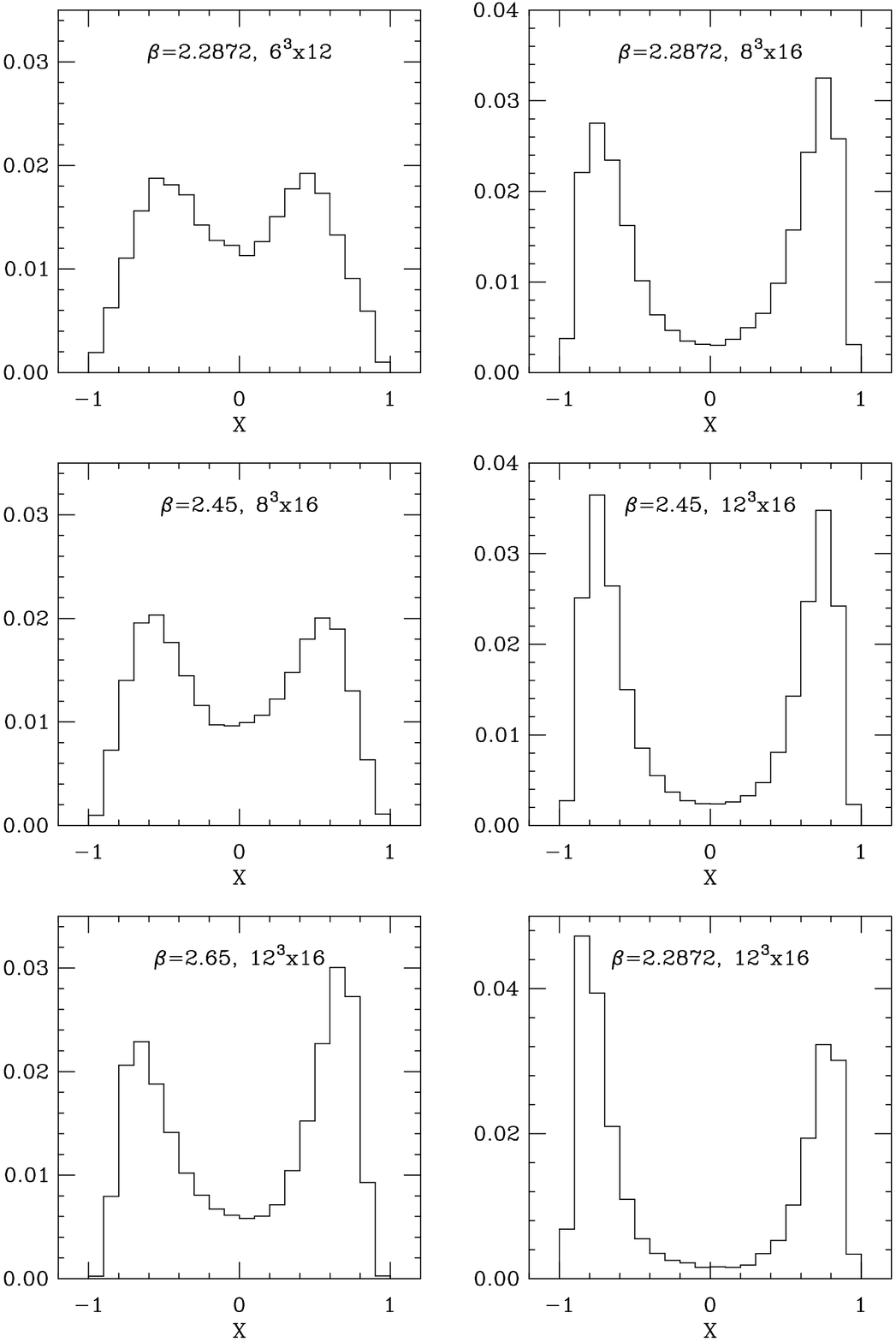}}}
\caption{Chirality histograms for the lowest two non-zero modes of the
overlap-Dirac operator at the 2.5\% sites with the largest
$\psi^\dagger \psi(x)$ on the six ensembles with Iwasaki gauge action.
The systems in the left column all have approximately the same volume in
physical units. The systems in the top two panels in the right column also
have the same, about a factor of 3 larger, volume, while the system in the
lower right hand corner panel has the largest volume.}
\label{fig:hist_IW_2l_2p5pc}
\end{figure}

The chirality histograms of the two lowest non-zero modes at the 2.5\%
of sites with largest $\psi^\dagger \psi(x)$ for all six ensemble with
Iwasaki glue are shown in Fig.~\ref{fig:hist_IW_2l_2p5pc}. The physical
volume of the systems shown in the left column are approximately the
same. Comparing the three histograms confirms the trend, already observed
with Wilson glue, that the lowest non-zero modes tend to become more
chiral in their peaks as the lattice spacing is decreased. This strongly
suggests that the behavior observed here will survive the continuum limit.
Comparing the three histograms for $\beta=2.2872$, on the other hand,
confirms the strong volume dependence already observed with Wilson glue.

\begin{figure}
\centerline{{\setlength{\epsfysize}{3in}\epsfbox[250 200 376 526]
 {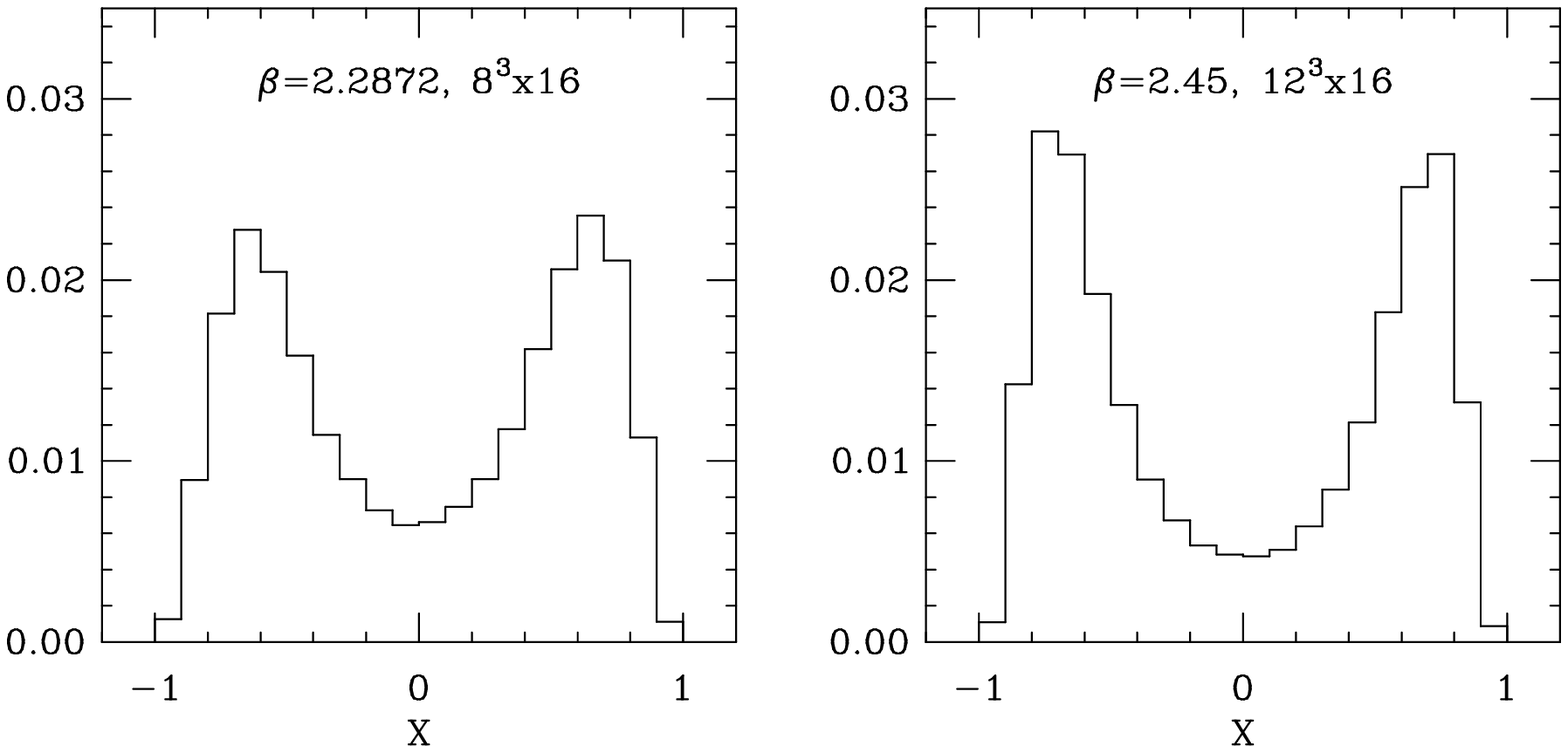}}}
\caption{Chirality histogram similar for the lowest six non-zero modes at
the 2.5\% sites with the largest $\psi^\dagger \psi(x)$ on the two lattices
with volume of about 7 fm$^4$.}
\label{fig:hist_IW_6l_2p5pc}
\end{figure}

\begin{figure}
\centerline{{\setlength{\epsfxsize}{4in}\epsfbox[0 0 576 576]
 {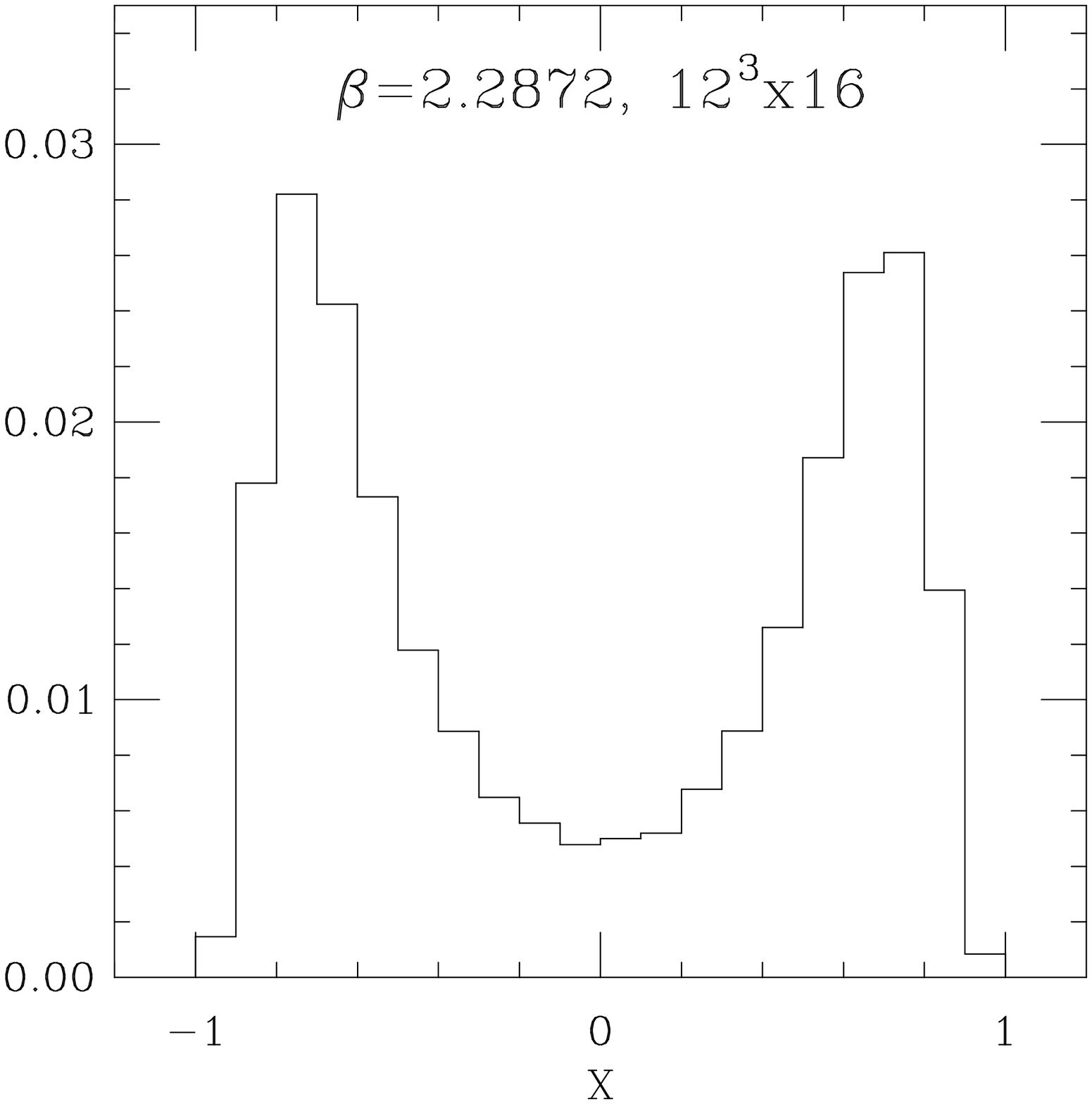}}}
\caption{Chirality histogram for all the non-zero modes out of the 20
lowest modes at the 2.5\% sites with the largest $\psi^\dagger \psi(x)$
on the lattice at $\beta=2.2872$ with the largest volume.}
\label{fig:hist_229LL_all_2p5pc}
\end{figure}

In Fig.~\ref{fig:hist_IW_6l_2p5pc} we show the chirality histogram of
the lowest six non-zero modes on the two systems with physical volume
of about 7 fm$^4$ and in Fig.~\ref{fig:hist_229LL_all_2p5pc} the chirality
histogram from all non-zero modes out of the 20 lowest modes that we
computed (there are between 11 and 20 non-zero modes per configuration).
Comparing with the histograms in the left column of 
Fig.~\ref{fig:hist_IW_2l_2p5pc} indicates that the number of non-zero
modes with similar chirality histograms grows roughly like the physical
volume so that there is a finite density of modes which are chiral in their
peaks. This is good evidence that we are not observing a finite volume
effect that would disappear in the infinite volume limit.

In the chirality histograms shown we typically considered the contribution
from the 2.5\% sites with the largest $\psi^\dagger \psi(x)$. In
Table~\ref{tab:psi_sq} we list the sum of $\psi^\dagger \psi(x)$ over
those sites as a percentage of the sum over all sites for some low-lying
eigenvectors for all the gauge field ensembles considered. The more chiral
the eigenvectors are on those 2.5\% sites, the higher is the percentage of the
eigenvector contained on those sites.

\section{Abelian gauge theories}

\begin{figure}
\centerline{{\setlength{\epsfxsize}{4in}\epsfbox[0 0 576 576]
 {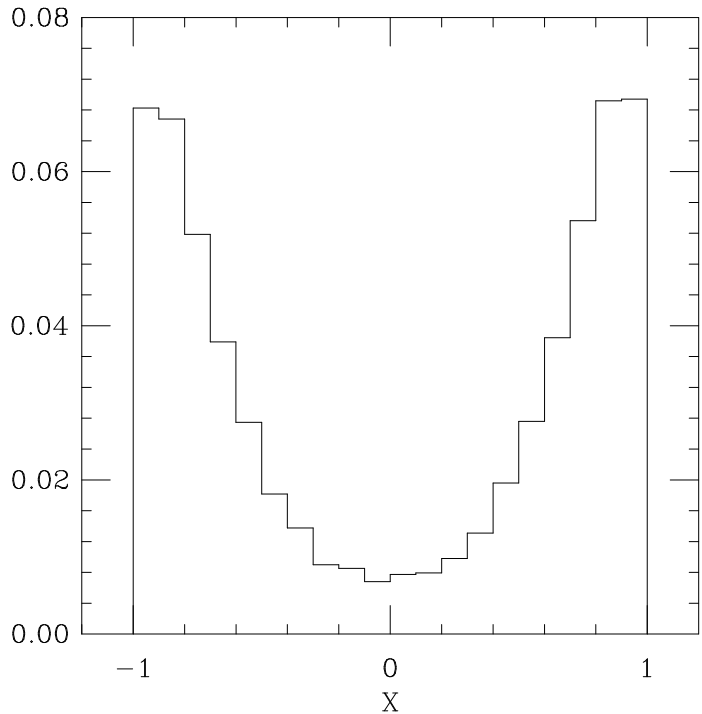}}}
\caption{Chirality histogram for the lowest ten non-zero modes of the 
overlap-Dirac operator at the 6\% sites with the largest 
$\psi^\dagger \psi(x)$ on $24^2$ U(1) configurations at $\beta=1.9894$.}
\label{fig:hist_u1d2_10l_6pc}
\end{figure}

To contrast the results for quenched QCD of the previous section we
considered the chirality parameter also for quenched U(1) theories in
two and four dimensions. In 2d, also considered in Ref.~\cite{HIMT},
topology plays a crucial role and one expects local chirality in the peaks
of low-lying modes. This is clearly seen in Fig.~\ref{fig:hist_u1d2_10l_6pc},
where the 10 lowest non-zero modes have been kept. Comparing with Fig.s~6
and 7 of Ref.~\cite{HIMT} demonstrates again the superiority of the
overlap-Dirac operator over the Wilson-Dirac operator for the purpose of
investigating
chirality properties. The histogram in Fig.~\ref{fig:hist_u1d2_10l_6pc}
for the 10 lowest non-zero modes shows better chirality peaks than
Fig.~6 of \cite{HIMT} for the real modes of the Wilson-Dirac operator,
the ``should be zero modes'', were it not for the explicit chiral symmetry
breaking from the Wilson term.

\begin{figure}
\centerline{{\setlength{\epsfysize}{3in}\epsfbox[250 200 376 526]
 {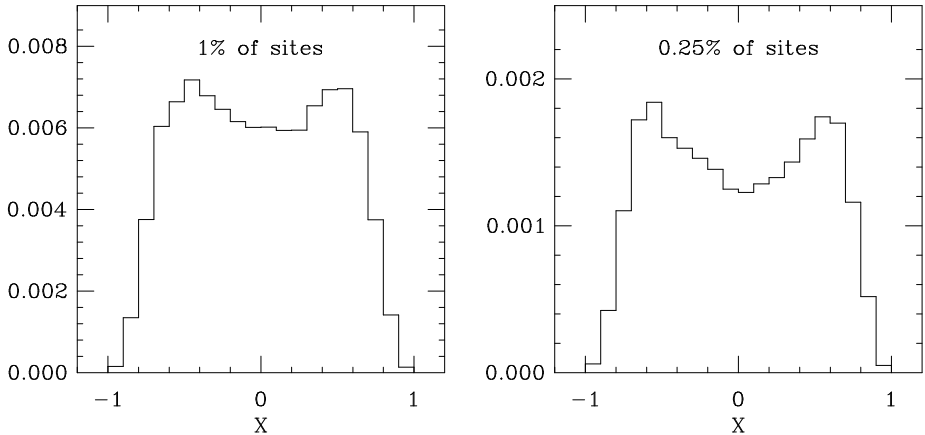}}}
\caption{Chirality histogram for the lowest two non-zero modes of the
overlap-Dirac operator at the 1\% (left panel) and 0.25\% (right panel)
sites with the largest $\psi^\dagger \psi(x)$ on $8^4$ U(1) configurations
at $\beta=0.9$.}
\label{fig:hist_u1d4_2l_comp}
\end{figure}

In 4d U(1), in the confined phase, chiral symmetry is spontaneously broken,
and a finite density of near zero modes exists. But there are no instantons,
and so one would not expect the near zero modes to be dramatically chiral even
in their peak regions. There do exist exact zero modes of the overlap-Dirac
operator, which again, of course, are chiral, although their origin is not
completely understood~\cite{U1_ov}. We analyzed, stored eigenmodes from
Ref.~\cite{U1_ov}. The chirality histogram for the lowest two non-zero modes
is shown in Fig.~\ref{fig:hist_u1d4_2l_comp}. We note that while there is
some mild indication of chirality peaking, the proportion of sites showing
this behavior is much reduced compared to the SU(3) case and appears to be
of a qualitatively different nature than seen before. 
Also, the sum of $\psi^\dagger \psi(x)$ over the sites kept for the
histograms of Fig.~\ref{fig:hist_u1d4_2l_comp} where almost identical
for all 12 lowest eigenmodes (4.0\% and 1.20\%, respectively; for the
2.5\% of sites with largest $\psi^\dagger \psi(x)$ the sum is about 8.4\%
for all eigenmodes as compared to the non-abelian case in
Table~\ref{tab:psi_sq}).
In the U(1) case the (near) lack of peaking could conceivably come about
from the scenario outlined by Horv\'ath, {\em et. al.} -- namely from the
confinement inducing vacuum fluctuations.

\section{Conclusions}

Horv\'ath {\em et al.}~\cite{HIMT} proposed an observable, the ``chirality''
parameter, designed to answer the question whether topological charge
fluctuations in (quenched) SU(3) gauge theory are instanton dominated
or not by the response of fermions.
Chirality of the low lying modes near $\pm 1$ in the regions where they
have the largest magnitude would signal instanton dominance.
Using Wilson fermions, afflicted by potentially strong lattice artifacts,
they claimed evidence that the topological charge fluctuations are {\it not}
instanton dominated. We took up this question using overlap fermions,
a lattice fermion formulation known to have the same chiral properties
as continuum fermions, and therefore much less afflicted by lattice
artifacts and better suited to the study of properties closely related
to chiral symmetry.

Considering several volumes and lattice spacings, we found convincing
evidence for chirality of the low lying modes in their peak region,
confirming results of Ref.~\cite{DH} obtained using overlap fermions
coupled to smoothened gauge fields.
We note that improving the gauge action did not significantly change
the degree of chirality at fixed lattice spacing and lattice size.
Our results give evidence that the
number of modes which are chiral in their peaks grows linearly with
the volume so that there is a {\it finite density} of such modes, and
that the chirality of the modes becomes more pronounced as the lattice
spacing is decreased. Therefore, our observations should remain valid in
the continuum limit. 

That the results of Ref.~\cite{HIMT} are strongly affected by lattice
artifacts was confirmed by the very recent paper of
Hip {\it et al.}~\cite{WC_imp}. These authors consider an improved
chirality parameter for Wilson fermions, clover improved Wilson fermions,
and lattices at smaller lattice spacing than Ref.~\cite{HIMT}. With
these reductions of lattice artifacts, they conclude that instanton
dominance of topological charge fluctuations is not ruled out by the
response of (improved) Wilson fermions. Together with the results from
overlap fermions presented in this paper and the lack of significant
chirality enhancement in the 4-d U(1) model where instantons should not
exist, the case for instanton domination in 4-d SU(3) gauge theory, 
as measured by the chirality parameter, becomes even more
compelling.


\section*{Acknowledgements}

RGE thanks Nathan Isgur and Hank Thacker for discussions. UMH thanks
Tom DeGrand and Anna Hasenfratz for discussions and Tom DeGrand for
suggesting this work.
RGE was supported by DOE contract DE-AC05-84ER40150 under which the
Southeastern Universities Research Association (SURA) operates the
Thomas Jefferson National Accelerator Facility (TJNAF).  UMH was
supported in part by DOE contract DE-FG02-97ER41022. The computations
reported here were performed on the JLab/CSIT QCDSP at JLab, 
the JLab/MIT workstation cluster and the CSIT workstation cluster.


\begin{thebibliography}{99}

\bibitem{inst_liquid} T. Sch\"afer and E. Shuryak, Rev. Mod. Phys.
{\bf 70}, 323 (1998).

\bibitem{diak} Cf. D. Diakonov, Lectures at the Enrico Fermi School in
Physics, Verenna 1995 [hep-ph/9602375].

\bibitem{Witten} E. Witten, Nucl. Phys. {\bf B149}, 285 (1979);
Nucl. Phys. {\bf B156}, 269 (1979).

\bibitem{HIMT} I. Horv\'ath, N. Isgur, J. McCune and H.B. Thacker,
hep-lat/0102003.

\bibitem{EHK_clov} R.G. Edwards, U.M. Heller and T.R. Klassen,
Phys. Rev. Lett. {\bf 80}, 3448 (1998).

\bibitem{DH} T. DeGrand and A. Hasenfratz, hep-lat/0103002.

\bibitem{TD_ov} T. DeGrand [MILC collaboration], Phys. Rev. {\bf D63},
034503 (2001).

\bibitem{Herbert} H. Neuberger, Phys. Lett. {\bf B417}, 141 (1998).

\bibitem{MILC_Pisa} C. Bernard {\it et al.} [MILC collaboration],
Nucl. Phys. {\bf B (Proc.Suppl.) 83-84}, 289 (2000).

\bibitem{DH_long} T. DeGrand and A. Hasenfratz, hep-lat/0012021.

\bibitem{EHN_practical} R.G. Edwards, U.M. Heller and R. Narayanan,
Nucl. Phys. {\bf B540}, 457 (1999).

\bibitem{ParalComp} R.G. Edwards, U.M. Heller and R. Narayanan,
Parallel Computing {\bf 25}, 1395 (1999).

\bibitem{Ritz} B. Bunk, K. Jansen, M. L\"uscher and H. Simma,
DESY-Report (September 1994); T. Kalkreuter and H. Simma, 
{\em Comput. Phys. Commun. \/} {\bf 93}, 33 (1996).

\bibitem{EHN_rho0} R.G. Edwards, U.M. Heller and R. Narayanan,
Nucl. Phys. {\bf B535}, 403 (1998).

\bibitem{DWF_CPPACS} CP-PACS Collaboration (A. Ali Khan, et. al.),
hep-lat/0007014.

\bibitem{DWF_eigs} C. Jung, R.G. Edwards, V. Gadiyak, and X.J. Xi
Phys. Rev. {\bf D63}, 054509 (2000).

\bibitem{DWF_BNL} 
T. Blum, P. Chen, N. Christ, C. Cristian, C. Dawson, G. Fleming, 
A. Kaehler, X. Liao, G. Liu, C. Malureanu, R. Mawhinney, S. Ohta, 
G. Siegert, A. Soni, C. Sui, P. Vranas, M. Wingate, L. Wu, Y. Zhestkov,
hep-lat/0007038.

\bibitem{DWF_IW} 
CP-PACS Collaboration (A. Ali Khan et al.)
Nucl. Phys. Proc. Suppl. {\bf 94}, 725 (2001).

\bibitem{PV} P. Vranas, Nucl. Phys. Proc. Suppl. {\bf 83}, 363 (2000).

\bibitem{Liu} K.-F. Liu, S.-J. Dong, F.X. Lee, and J.B. Zhang,
Nucl. Phys. Proc. Suppl. {\bf 94}, 752 (2001).

\bibitem{EHK_string}
R.G. Edwards, U.M. Heller, T.R. Klassen,
Nucl. Phys. {\bf B517}, 377 (1998).

\bibitem{IW} Y. Iwasaki, preprint UTHEP-118 (1983), unpublished;
Y. Iwasaki and T.Yoshie, Phys. Lett. {\bf 131B}, 159 (1983).

\bibitem{LW}
See M. L\"uscher and P. Weisz, 
Phys. Lett. {\bf 158B}, 250 (1985), and references therein;
M. Alford, W. Dimm, G.P. Lepage,
Phys. Lett. {\bf B361}, (1995) 87.

\bibitem{U1_ov} B.A. Berg, U.M. Heller, H. Markum, R. Pullirsch and
W. Sakuler, hep-lat/0103022.

\bibitem{WC_imp} I. Hip, Th. Lippert, H. Neff, K. Schilling and W. Schroers,
hep-lat/0105001.

\end{thebibliography}
\end{document}